\documentclass[a4paper,x,,12pt]{article}
\usepackage[round]{natbib}
\usepackage{amsopn}
\usepackage[it]{caption}
\usepackage{float}

\usepackage[T1]{fontenc}
\usepackage{times}
\usepackage{graphicx}
\usepackage{epstopdf}
\usepackage{url}
\usepackage[]{algorithm2e}
\usepackage{hyphenat}
\usepackage{amsmath}
\usepackage{mathtools}
\usepackage{setspace}

\usepackage[left]{lineno}

\begin{document}

\begin{flushright} version 261030  \end{flushright}
\vspace {2 cm}

\noindent {\bf \LARGE \center Feed-forward Active Magnetic Shielding}

\vspace{0.5 cm}

\noindent Alain de Cheveign\'e (1, 2, 3)
\vspace{2 cm}

\vspace{2 cm}


\noindent AUTHOR AFFILIATIONS:

\noindent (1) Laboratoire des Syst\`emes Perceptifs, UMR 8248, CNRS, France.

\noindent (2) D\'epartement d'\'Etudes Cognitives, Ecole Normale Sup\'erieure, PSL, France.

\noindent (3)  Department of Psychology, University College London.

\vspace{2 cm}

\noindent CORRESPONDING AUTHOR:

\noindent Alain de Cheveign\'e, Audition, DEC, ENS, 29 rue d'Ulm, 75230, Paris, France, Alain.de.Cheveigne@ens.fr, phone 00447912504027.
\vspace{2 cm}

\newpage

\section*{Abstract}

Measurement of small magnetic fields requires shielding. For example,  magnetoencephalography (MEG) measures small fields arising within the brain, in the presence of environmental fields several orders of magnitude larger.  A common solution is passive shielding, but its cost, bulk and limited effectiveness motivate the alternative approach of active shielding, which comes in two flavours: feed-back and feed-forward.  In feed-back shielding (the most common design), coils are driven from sensors that measure the corrected field. In feed-forward shielding (less common), coils are driven from sensors that measure the uncorrected field.  The latter has the potential of providing better performance, but its implementation  is hobbled by an inevitable  coupling between coils and reference sensors.  This paper describes an approach  that employs  a data-driven ``decoupling matrix" to suppress the coupling, enabling greater flexibility to achieve effective shielding across a wide frequency range. The method is evaluated by simulation, first using a simplified "flat world" in which the principle is clearly visualised, and then based on the more realistic geometry of a standard  MEG system. Performance appears to be limited only by spatial sampling and numerical resolution. The method, which  is data-driven and does not require geometric calculations, tight-tolerance fabrication, or manual calibration, is  complementary with standard feed-back shielding methods. Active shielding is of particular interest in the context of a new generation of optically-pumped magnetometers (OPM), as the lower cost and reduced constraints of these new systems relative to SQUID-based systems provide an incentive to likewise reduce the cost and constraints of shielding.

\paragraph{Keywords:}\ magnetic shielding, active magnetic shielding, magnetoencephalography (MEG), feed-back, feed-forward, magnetometer, coil

\newpage.

\section{Introduction}

\begin{figure}
\includegraphics[scale=.45]{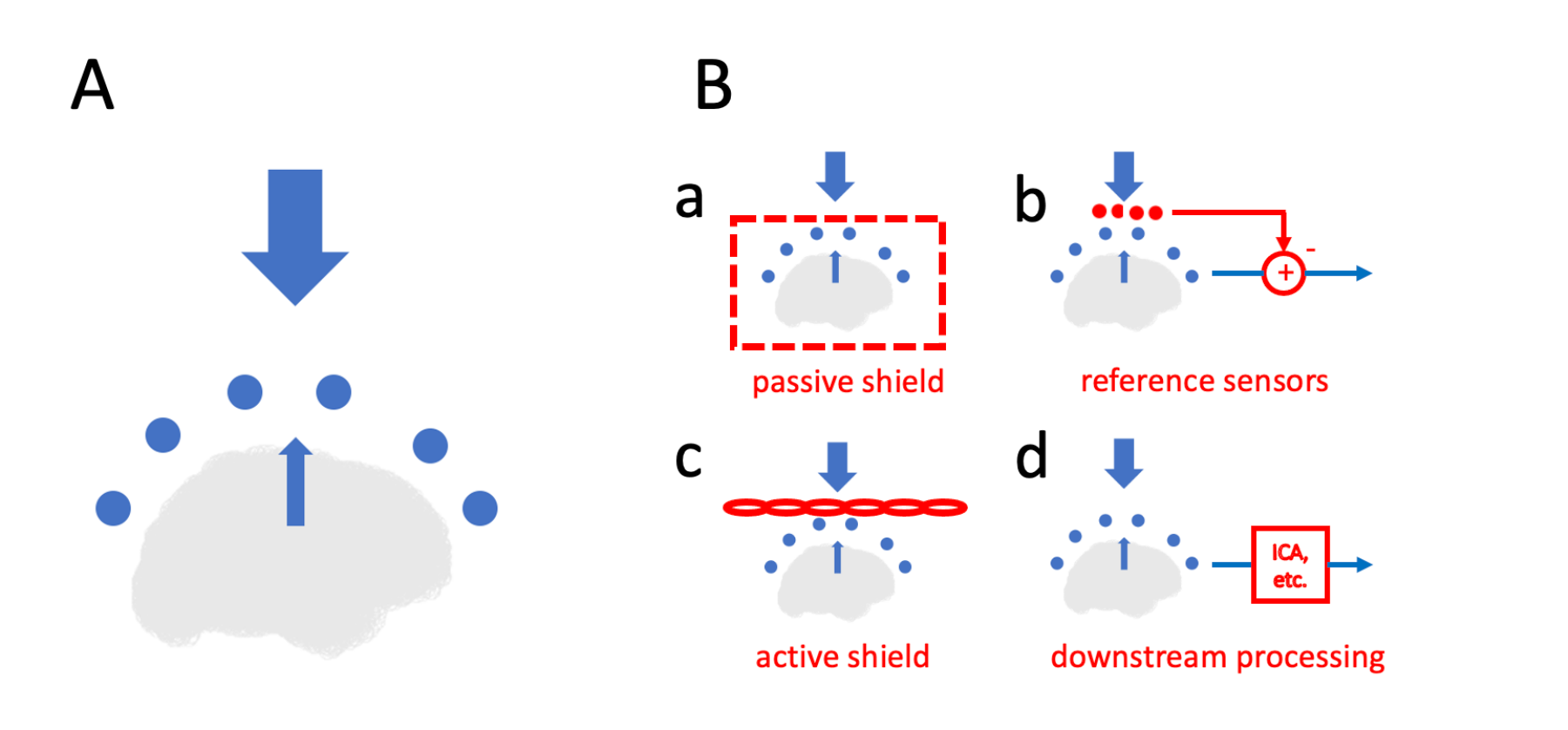}
\caption{\label{problem}
A: Magnetic fields from the brain are dwarfed by fields from ambient sources, including the earth's magnetic field and fields from electrical circuits and machinery. B: Classic approaches to the problem: passive magnetic shielding, subtraction of signals from reference sensors, active magnetic shielding with coils, downstream signal processing. 
 }
\end{figure}

Bio-magnetic fields from the brain are much smaller than ambient magnetic noise fields from vehicles, machines or power lines, or the  earth's magnetic field (Fig.~\ref{problem}A). Suppressing  ambient fields is a daunting task because of their magnitude:  up to 10$^{9}$ (10$^{18}$ in power)  times those of brain fields of interest \citep{weinstock_multichannel_2000}.  Large fields  swamp  smaller fields, eat up dynamic range,  make magnetic measurements  sensitive to  structural vibration, and prevent certain devices such as optically-pumped magnetometers (OPM) from functioning. This paper describes a new approach to active magnetic shielding that complements existing approaches and expands the solution space. 

The problem can be tackled in several ways that are sometimes used in combination \citep{seymour_interference_2022}. The most straightforward is {\em passive shielding}  by layers of mu-metal that mainly attenuate low frequency fields, and aluminium or copper that attenuate high frequencies (Fig.~\ref{problem}Ba). Drawbacks are cost and weight, limited usable space, a limited attenuation factor (10$^2$ to 10$^4$) \citep{brake_improvement_1991, weinstock_multichannel_2000}, and the need to de-gauss the shield in certain circumstances.

Another approach is to replace magnetometers by {\em gradiometers}\ more sensitive to nearby brain sources than distant ambient sources. 
A gradiometer works by subtracting  the field  at a secondary coil offset in space, and normalizing the difference by the distance.  The secondary coil can be replaced by an array of {\em reference sensors} (Fig.~\ref{problem}Bb), the output of which is scaled and subtracted in the digital domain to create ``synthetic gradiometers'' that can be of higher order and fine-tuned, with  coefficients found either analytically, based on the geometry,  or by data-driven techniques such as regression \citep{weinstock_multichannel_2000, de_cheveigne_denoising_2007}. 

{\em Downstream processing}\  in the digital domain (Fig.~\ref{problem}Bd) relies on differences between ambient and target fields in the {\em spatial} or {\em spectral} domains, using spatial or spectral filters. Here too, there is a choice between pre-determined solutions (e.g.\ gradient, Laplacian or spherical harmonic decomposition), and data-driven solutions  for which a vast toolbox of linear data-driven methods is available \citep{Parra2005}.  Downstream processing requires a wide dynamic range for the sensors, the analog-to-digital converters, and the numerical representation used for computation \citep{seymour_interference_2022}. 

A final approach is {\em active shielding}\ (Fig.~\ref{problem}Bc) in which ambient fields are cancelled by correction coils fed with a combination of fixed current (e.g., to cancel the earth's field) and time-varying current (to cancel the fluctuating components of the  field) \citep{cohen_large-volume_1970, xu_active_2026, Bezsudnova2026.08.06.743270}.  Suppressing the field before transduction, rather than after,  reduces dynamic range requirements on sensors and sensitivity to vibration. Careful design of the coils ensures a uniform corrective field (to null a uniform ambient field) \citep{holmes_bi-planar_2018}, or instead an appropriately ``sculpted'' corrective field (to null a non-uniform ambient field) \citep{holmes_balanced_2019,iivanainen_scalp_2019, zheng_design_2025,tian_design_2025}. 

For a fixed field, the currents that drive the coils might be adjusted manually, but for a time-varying field  they must be guided by sensors that track that field. Multiple ambient sources require multiple sensors to drive  multiple coils via a sensor-to-coil driving matrix ${\bf M}$ or controller. Some applications (e.g. MEG with a fixed geometry) require cancellation at the primary sensors, others (e.g., a physics experiment or industrial process)  require cancelling within a  volume. For yet others (e.g. OPM-based MEG with a moving subject), the driving matrix ${\bf M}$ might need to vary in time.

%
\begin{figure}
\includegraphics[scale=.3]{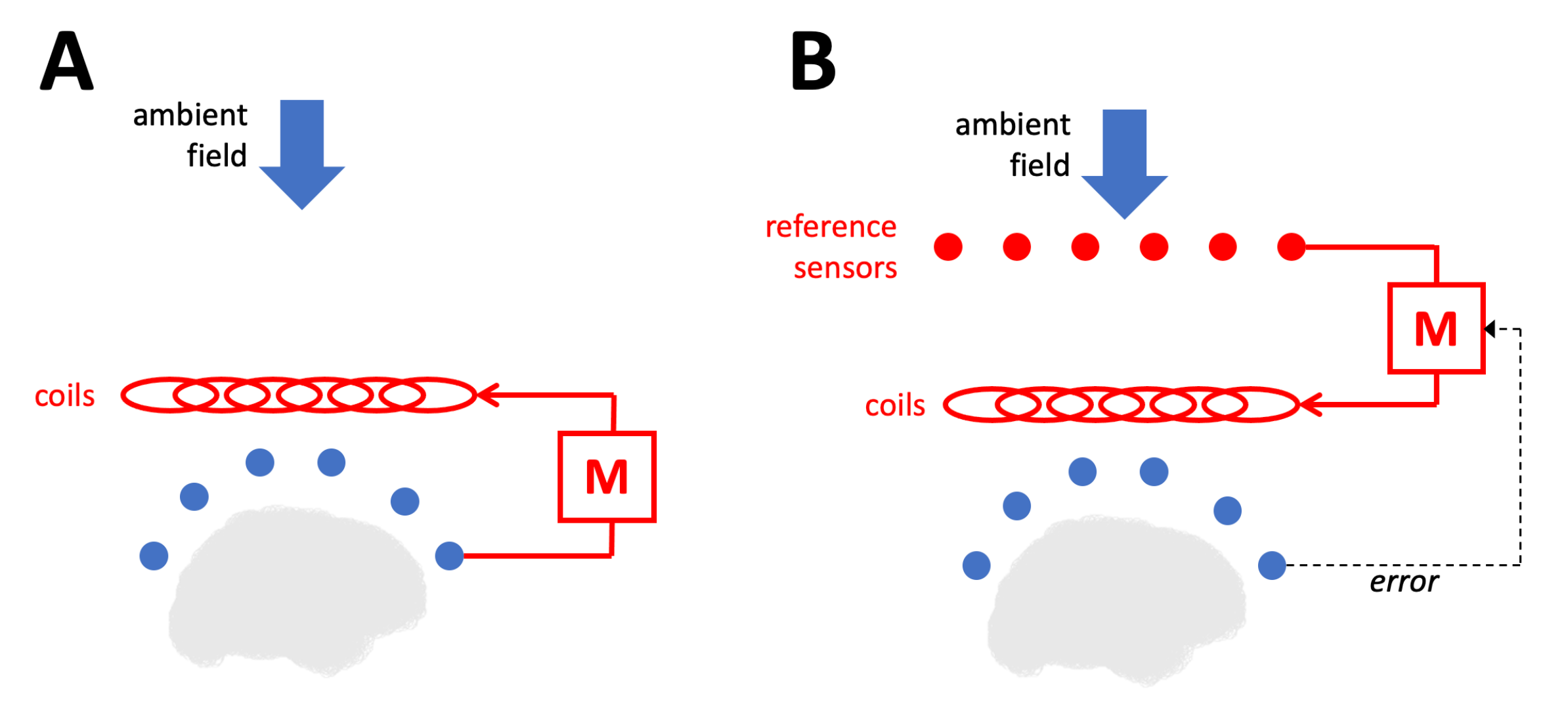}
\caption{\label{topology}
Two basic topologies for active shielding of an MEG system. A: Feed-back: coils are driven from the primary sensors via a matrix {\bf M} tuned to minimise the field at the primary sensors. B: Feed-forward: coils are driven from an array of reference sensors via a matrix {\bf M} tuned to minimise the field at the primary sensors.
 }
\end{figure}

Active shielding methods can be divided into two classes according to whether the coils are driven by the sensors to be shielded (closed-loop or feed-back topology, Fig.\ \ref{topology}, left), or by a separate set of reference sensors  (open loop or feed-forward topology, Fig.\ \ref{topology}, right). 
The feed-back topology is a popular choice   \citep[e.g.][see Discussion]{iivanainen_scalp_2019, Bezsudnova2026.08.06.743270}. Typically, a PID (proportional-integral-differential) controller minimises the field at a sensor subject to disturbance from the ambient field. However, feed-back shielding is vulnerable to a number of issues.
\begin{enumerate}
\item 
If cancellation were perfect, there would be no signal to drive the coils (ascending path in Fig.\ \ref{topology}, left). Thus it cannot be perfect.
\item
Cancelling of the field at a sensor prevents measuring a field at that sensor \citep{volegov_noise-free_2004}. 
\item
The parameters of a PID controller require tuning, with a tradeoff between performance and stability.
\item
If there are multiple coils and sensors, their control paths may interact, requiring a more complex multi-input-multi-output (MIMO) design.
\end{enumerate}.
These issues can be dealt with, witness many examples of successful shielding  \citep[e.g.][see Discussion for more details]{schofield_novel_2024, xu_active_2026}. 
However,  the added complexity and constraints  motivate  alternative approaches, possibly complementary to these, to facilitate design and improve performance.

The feed-forward topology  (Fig. \ref{topology}, right),  less common \citep[e.g.][see Discussion]{pyragius_high_2021}, infers the ambient field at the primary sensors from measurements at the reference sensors, avoiding some of the issues just enumerated. In particular the absence of a loop in the data pathway avoids the risk of instability. However, the presence of a spurious coupling between coils and reference sensors, hard to rule out, blurs the distinction and weakens this advantages.  The decoupling matrix method described in this paper removes this coupling,  simplifying the design and  allowing better shielding performance.

This new approach borrows ideas from linear data analysis \citep[e.g.,][and references therein]{Parra2005} and sound and vibration control \citep{cabell_principal_1998, poshtan_multichannel_2003}. The goal is to null the ambient magnetic field at an array of primary sensors for the purpose of magnetic measurement (as in MEG), Additional goals such as shielding a {\em volume}\ (e.g., for an industrial process or physics experiment) or dealing with a {\em time-varying}\ configuration (as in ambulatory MEG) are alluded to, but not addressed in detail.

\section{Methods} 
\label{methods}

The method assumes an array of primary sensors, an array of reference sensors, an array of correction coils, and electronics that implement the necessary linear algebra operations and produce driving currents.  It operates in two modes: {\em learning}\ (data-driven calibration), and {\em shielding}. 

\subsection{Data model}

The primary sensor signal matrix ${\bf x}(t)$ is of size $J\times T$ where $T$ is the number of samples within a time interval and $J$ is the number of primary sensors, and the reference sensor matrix ${\bf r}(t)$ is of size $K\times T$ where $K$ is the number of reference sensors. Corrective currents produced by the system are denoted by ${\bf q}(t)$ of size $L\times T$ where $L$ is the number of coils (Fig.\ 3a). The ambient field is assumed to come from multiple sources represented by a matrix ${\bf s}(t)$ of size $I \times T$ where $I$ is the number of sources.  Data matrices ${\bf x}(t)$ and ${\bf r}(t)$ are measured, ${\bf q}(t)$ is produced, and ${\bf s}(t)$ is unknown. 

The field at the primary sensors includes contributions from both ambient sources and coils:
\begin{eqnarray}
{\bf x}(t)&=&{\bf As}(t)+{\bf Cq}(t) 
\end{eqnarray}
where ${\bf A}$ and ${\bf C}$ are source-to-primary-sensor gain matrices (Fig.\ 3b). Likewise:
\begin{eqnarray}
{\bf r}(t)&=&{\bf Bs}(t)+{\bf Dq}(t)
\end{eqnarray}
where ${\bf B}$ and ${\bf D}$ are source-to-reference-sensor gain matrices (Fig.\ 3b). The coil-to-primary-sensor gain matrix ${\bf C}$ can be calculated based on Maxwell's equations if the geometry is known, however it may be easier to estimate it by driving the coils with a current ${\bf q}_d$(t) large enough so that the ambient field can be ignored and calculating: 
\begin{eqnarray}
{\bf C}&=&{\bf {\bf x}_d (t) q}_d(t)^\dag\label{D}
\end{eqnarray}
where  ${\bf x}_d (t)$ and ${\bf r}_d (t)$ are sensor data and ${\bf q}_d (t)^\dag$ is the pseudo inverse of the drive current ${\bf q}_d (t)$ ($d$ indicates {\it drive}); the matrix  ${\bf D}$ could be estimated in the same way as ${\bf C}$, but this is not required.
The ambient-source-to-sensor matrices  ${\bf A}$ and  ${\bf B}$ cannot be estimated, but we can define a ``predictor matrix''  ${\bf P}$ that predicts (or infers) the ambient field at the primary sensors from its values at the reference sensors.  ${\bf P}$ is estimated, in the learning phase, from data recorded in response to the ambient field in  the absence of a drive current: 
\begin{eqnarray}
{\bf P}&=&{\bf x}_a (t){\bf r}_a (t)^\dag \label{P}
\end{eqnarray}
($a$ indicates {\em ambient}).
For this estimate to be reliable, ambient sources must all be active during the learning phase.
In the shielding phase, the coils are driven from the reference sensors as
\begin{eqnarray}
{\bf q}(t)&=&{\bf Mr}(t) \label{Q}
\end{eqnarray}
where ${\bf M}$ of size $L\times K$ is a reference-to-coil gain matrix (Fig.\ 3A) calculated in the learning phase as described below.

\begin{figure}
\includegraphics[scale=.35]{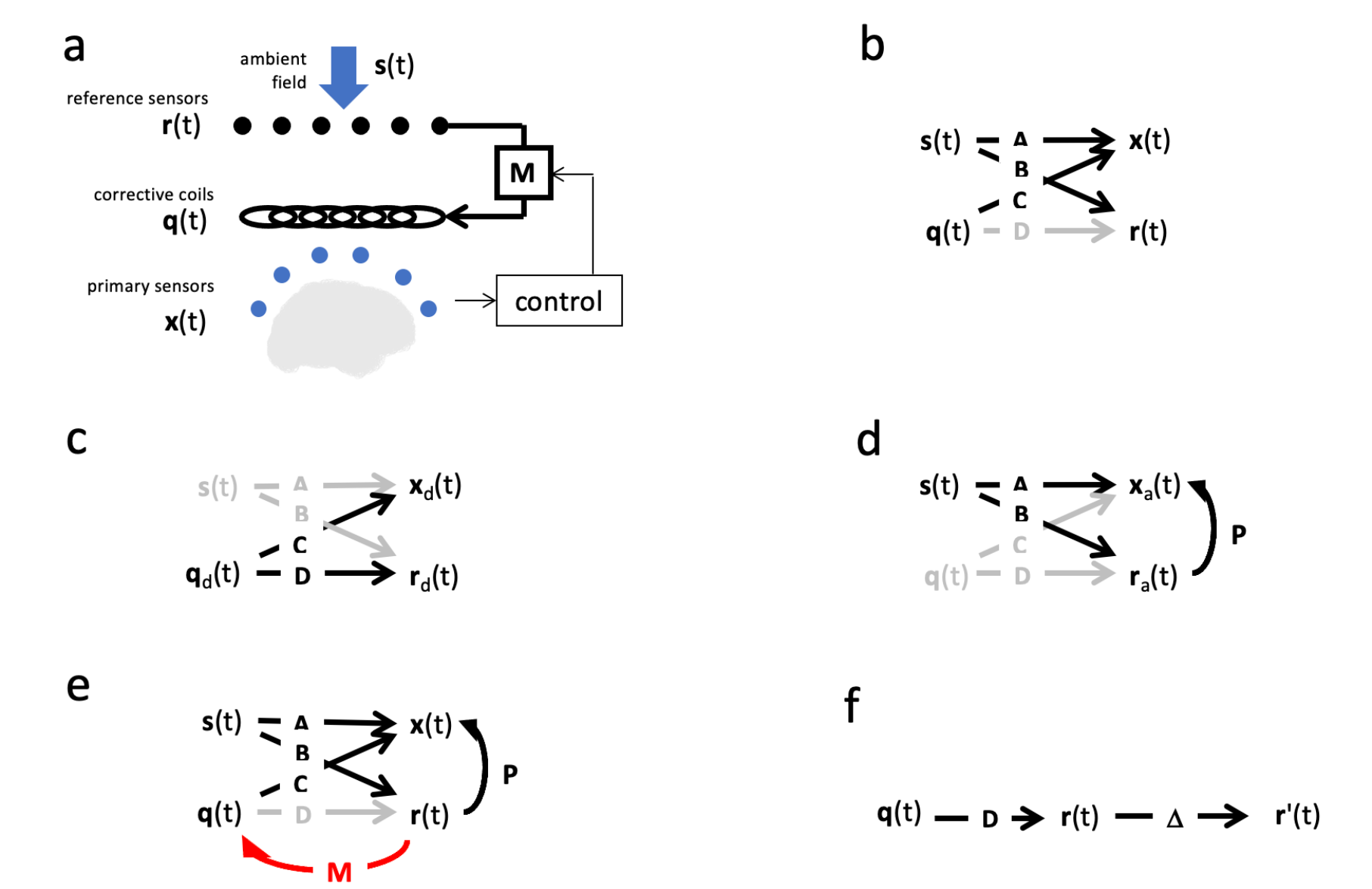} 
\includegraphics[scale=.35]{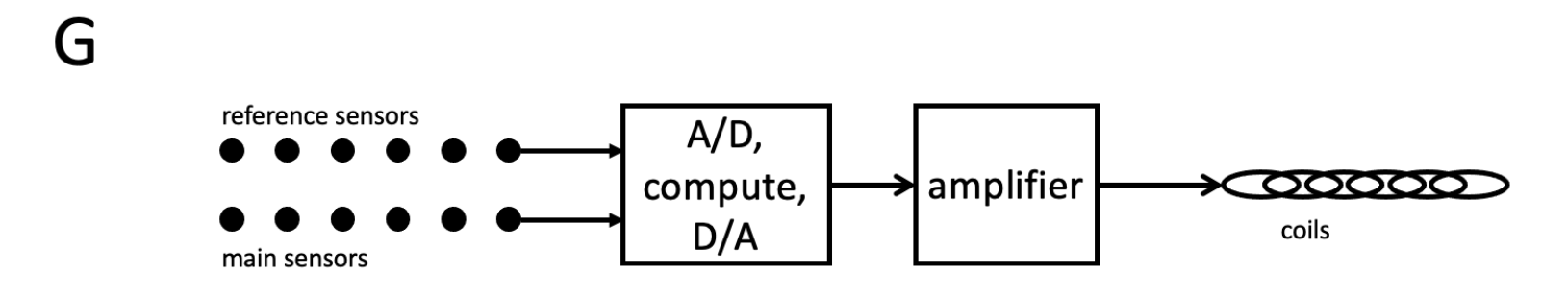}
\caption{\label{method}
Principle of the feed-forward active-shielding method. a: An array of correction coils is driven from an array of reference sensors via a feed-forward sensor-to-coil transform matrix  ${\bf M}$ that is designed to minimise the ambient field at the primary sensors. b: Block diagram of data and transform matrices.   c: Coil-to-sensor matrices  ${\bf C}$  is estimated by driving coils with a current  ${\bf q}_d (t)$ large enough that the ambient field can be ignored. d: The reference-to-primary sensor predictor matrix  ${\bf P}$ is estimated by recording sensor outputs in response to the ambient field, in the absence of coil current.  e: If the coil-to-reference coupling matrix  ${\bf D}$ is small enough be ignored,  ${\bf M}$ can be estimated from  ${\bf C}$  and  ${\bf P}$ as shown.  f: If coupling matrix ${\bf D}$ is non-negligible, its effect can be removed by applying a ''shielding'' matrix  ${\bf \Delta}$ (see text).
G: Structure of the hardware system.
}
\end{figure}

\subsection{Finding ${\bf M}$} \label{findingM}

If the coil-to-reference sensor coupling ${\bf D}$ were negligible (which might be approximately true if the coils were far from the reference sensors), the  matrix ${\bf M}$ could be calculated as 
\begin{eqnarray}
{\bf M}& = &-{\bf C}^\dag {\bf P}  \label{M}
\end{eqnarray}  
but if ${\bf D}$ is not negligible that solution is inaccurate. 
One might attempt to apply a corrective factor
\begin{eqnarray}
( {\bf I}-{\bf D}{\bf C}^\dag {\bf P})^\dag  \label{corrective_factor}
\end{eqnarray} to compensate for the coupling, however it is applicable only if the largest eigenvalue of ${\bf D}{\bf C}^\dag {\bf P}$ is less than one, otherwise the system is unstable.  Spurious coil-to-reference sensor coupling is a basic vulnerability of feed-forward active shielding.
 A different approach, used here, is to define a  matrix ${\bf \Delta}$ such that  ${\bf \Delta D} = 0$ and apply this ``decoupling'' matrix   to ${\bf r}(t)$ to suppress the effects of the coupling (Fig.~\ref{method}f). 
A suitable matrix ${\bf \Delta}$  exists as long as the number of reference sensors exceeds the number of  coils by a sufficient margin. 
It can be found by driving the coils with a current matrix ${\bf q}_d(t)$ of full rank, and applying PCA to ${\bf r}_d (t)$.
Since ${\bf r}_d (t)$ is of rank at most $L$ (number of coils), the last $K-L$ rows of the PCA matrix project ${\bf q}_d (t)$ to zero, and thus
populating ${\bf \Delta}$ with these  rows yields the desired decoupling matrix.  Thanks to this matrix, Eq.\ \ref{M} can be applied with no risk of instability, and the full  benefits of feed-forward active shielding are achieved.
To summarise, the three steps of the learning phase are:
\begin{enumerate}
\item Drive the coils with a current matrix ${\bf q}_d$ (t) of full rank and amplitude sufficient so that ambient noise can be ignored, record primary and reference sensor matrices ${\bf x}_d (t)$ and ${\bf r}_d (t)$, and derive ${\bf C}$ and ${\bf \Delta}$.
\item In the absence of a driving current, record ${\bf x}_a (t)$ and ${\bf r}_a (t)$, and derive ${\bf P}$.
\item Calculate ${\bf M}$.
\end{enumerate}


\section{Results}
The method is simulated first within a simplified two-dimensional  ``flat world'' that allows for easy visualisation, and then within a more realistic 3D configuration. 

\subsection{Flat world simulation}

To gain insight, a ``flat world'' is defined in which sensors and sources are located within a plane, each source producing a scalar ``field'' that varies isotropically with distance as $1/d^2$.  This  captures the structure of the problem while allowing for easy visualisation: complexities related to 3D positions and orientations of sources and sensors are conveniently ignored.

\begin{figure}
\includegraphics[scale=.55]{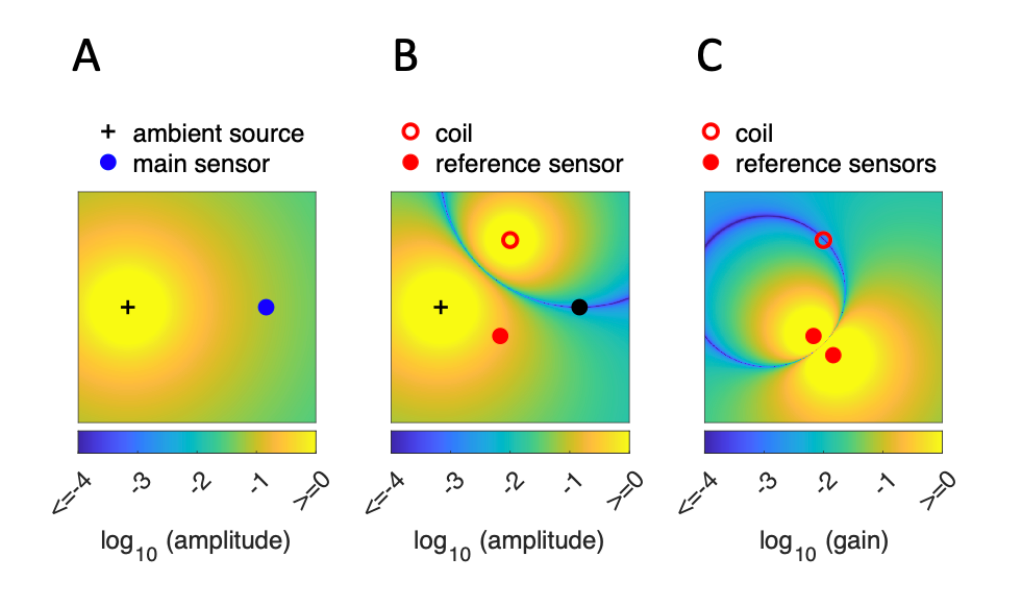}
\caption{\label{toy}
``Flat world'' simulation. A: The amplitude of the field produced by an ambient source (black cross) is represented as colour. The black dot represents the location of the primary sensor.  B: After adding the corrective field, produced by the coil (red circle) driven from the reference sensor (red dot), the corrected field (represented as colour) is zero at the primary sensor. C: Gain pattern of a linear combination of two reference sensors with weights tuned to minimise gain from the coil (``decoupling matrix'').}
\end{figure}

Figure\ \ref{toy}A represents as colour the field produced by one ambient source (cross). Its amplitude, which decreases with distance, is not zero at the primary sensor (dot): if that sensor were intended to measure the field of a weak source (not shown), the measurement would be contaminated.  Panel B shows the result of adding a corrective field produced by a coil (red circle) driven from a reference sensor (red dot) that tracks the ambient field. With appropriate gain, the ambient field is cancelled at the primary sensor (blue line). This is the principle of  feed-forward active shielding.
Unfortunately, the reference sensor also picks up the corrective field, creating a loop that makes it harder to calculate the required parameters and may introduce instability. This is avoided by using two reference sensors (red dots in panel C) that are combined so that the coil-to-reference-sensor-array gain (represented as colour) is zero at the coil, ensuring a corrective field that is easy to calculate, accurate and reliable.

\begin{figure}
\includegraphics[scale=.55]{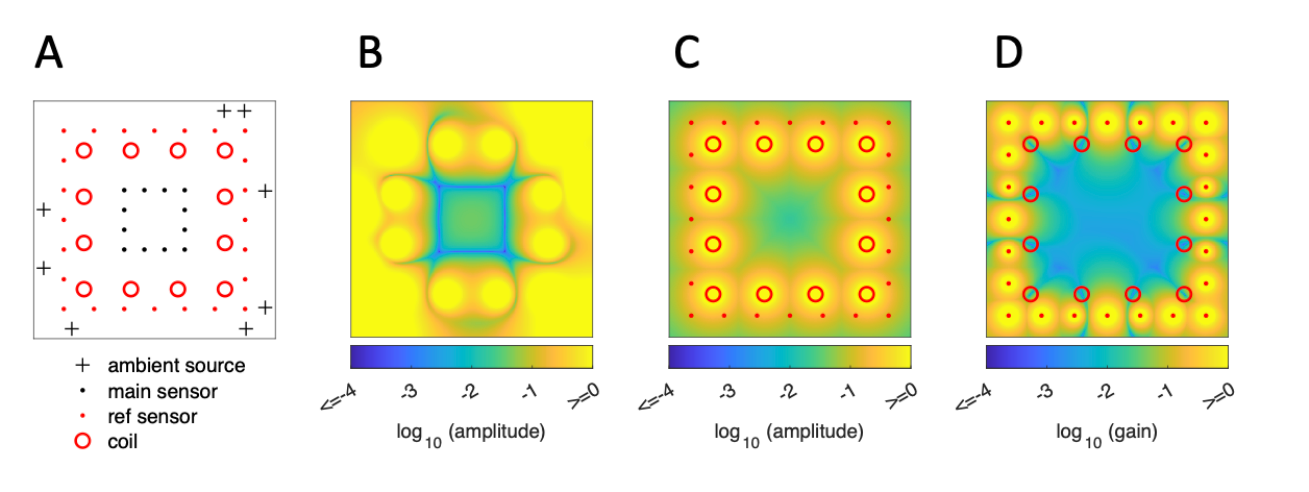}
\caption{\label{toy2}
Flat world simulation, complex configuration. A: Locations of ambient sources (black crosses), primary sensors (black dots) coils (red circles), reference sensors (red dots).  B: RMS amplitude of the ambient field after feed-forward shielding with decoupling matrix (note the small values, blue, at primary sensor positions). 
C: RMS amplitude of the field created by the coils (supposing equal amplitudes). D: RMS gain pattern of the reference sensors after application of the decoupling matrix.  Note the very small values at the coils (tiny blue dots centered on coils).}
\end{figure}

Figure \ref{toy2} illustrates a more complex simulation. Locations of sources and sensors are shown in panel A.  In the absence of shielding, the ambient source fields extend throughout the plane; in the presence of feed-forward active shielding with a decoupling matrix (panel B), the ambient field is effectively suppressed at the primary  sensors (dark blue). Key to this result is the decoupling matrix to counter the very strong coupling between coils and reference sensors (panel C shows the amplitude of the field that the coils would produce if driven with equal currents). The effect of the decoupling matrix is to  put local zeros at the positions of the coils (tiny blue dots centered on each coil in panel D), while maintaining non-zero gain at the ambient source locations. With decoupling, the reference sensors ``see'' the latter, but not the former hence no loop.

To implement the simulation that produced Fig.\ \ref{toy2}, the inverse square law was applied to obtain ground-truth source-to-sensor gain matrices ${\bf A}_0$, ${\bf B}_0$, ${\bf C}_0$, ${\bf D}_0$, characteristic of the ``physics'' of the system. These matrices are not accessible to the learning algorithm. 
In a first step of the learning phase, the coils were driven with a 12$\times$1000 matrix ${\bf q}_d (t)$ of Gaussian random numbers, and the primary and reference sensor matrices were obtained as ${\bf x}_d (t)={\bf C}_0{\bf q}_d (t)$ and ${\bf r}_d (t)={\bf D}_0{\bf q}_d (t)$. Based on these ``measurements,'' the coil-to-primary sensor gain matrix was estimated as described in the Methods, and the decoupling matrix by applying PCA to ${\bf r}_d (t)$ and populating ${\bf \Delta}$ with the last 12 rows of the 24$\times$24 PCA matrix. 
In a second step, ambient source activity was simulated as a 8$\times$1000 matrix ${\bf s}_a (t)$ of Gaussian-distributed random numbers and the primary and reference sensor  matrices were obtained as ${\bf x}_a (t)={\bf A}_0 {\bf s}_a (t)$ and ${\bf r}_a (t)={\bf B}_0{\bf s}_a (t)$, and the reference-to-primary sensor predictor matrix and reference-to-coil gain  matrix  calculated as explained in the Methods. 

For the shielding phase, an ``ambient'' source matrix ${\bf s} (t)$ was simulated as an 8-row matrix of Gaussian-distributed random numbers,  reference sensor signals were obtained as  ${\bf s}(t)={\bf B}_0{\bf s}(t)$, and the coil-driving currents were produced as  ${\bf q}(t)={\bf M\Delta r}(t)$.  Multiplying this by ${\bf C}_0$ produced the corrective field that, subtracted from the ambient field, yielded the corrected field (panel B).

\subsection{Simulation  in 3D space} \label{3D}

Whereas the previous simulations assumed idealized ``sources'' and ``sensors" within a plane with a simplified  $1/d^2$ law for ease of visualization,  here, the simulations assume realistic sensors and sources distributed in 3D space, with fields that obey the Biot-Savart law.

\paragraph{Minimal configuration.}

\begin{figure}
\includegraphics[scale=.4]{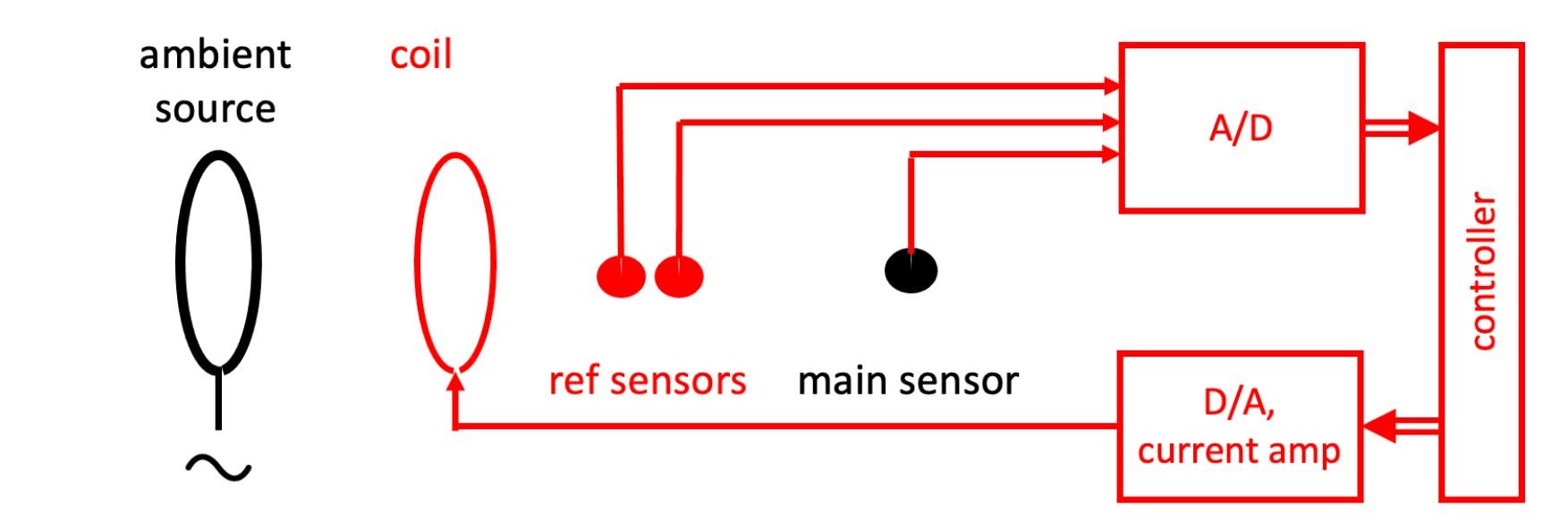}
\caption{\label{poc}
Minimal configuration: one ambient source, one coil, two reference magnetometers and one primary magnetometer, together with A/D and D/A converters, a current amplifier, and a 'controller' computer running real-time software. 
}
\end{figure}

Figure \ref{poc} depicts the simplest physical configuration that allows for ``proof of concept'' of feed-forward cancellation with a decoupling matrix.  In addition to an ambient field source and the primary magnetometer to be protected (in black), it includes one coil, two reference magnetometers, and the associated electronics (in red). Source and sensor orientations are aligned, the objective being to control the field along that direction.
%

\paragraph{Realistic MEG configuration.}

\begin{figure}
\includegraphics[scale=.5]{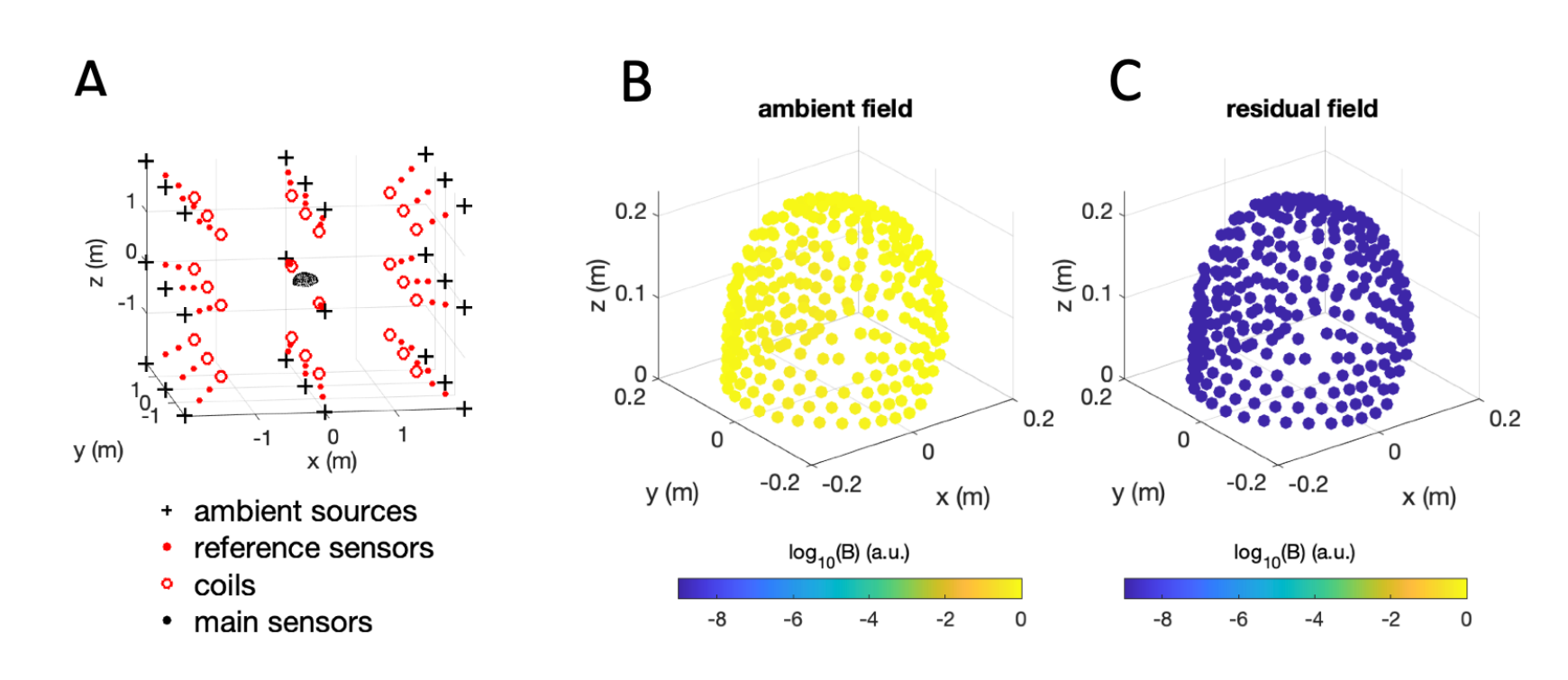}
\caption{\label{simulation}
Simulation in 3D space based on a real MEG system. A: Locations of ambient sources (black crosses), primary sensors (small black dots near the center), reference sensors (red dots) and coils (red circles) within a cube of side 4m.  The primary sensors are located and oriented as in a CTF MEG system.  Ambient sources and coils are modelled as dipoles.  Ambient sources, coils and reference sensors have arbitrary positions and randomly-chosen orientations.  B: RMS magnitude of ambient field over primary sensors (arbitrary units). C: same, after ambient field suppression.
}
\end{figure}

This configuration is centered on an array of 274 magnetometers with positions and orientation chosen to match those of the array of gradiometers in a CTF MEG system (Fig.~\ref{simulation}A).  This sensor array was placed within an ambient field created by a set of 26  sources  simulated as randomly-oriented dipoles on a cube of side 4~m centred on the CTF coordinate origin. 
A feed-forward active shield was implemented with an array of 26 randomly-oriented coils distributed on a cube of size 2.8~m, and an array of 52 randomly-oriented reference sensors distributed on two cubes of size 3.2~m and 3.4~m respectively.   
The number and geometry of ambient sources, coils and reference sensors is arbitrary, the intent being to simulate a challenging situation. For computational convenience, coils were modelled as dipoles which, if anything, increases the challenge because the fields they produce are less uniform than expected from Helmholtz coils.

The ``physics'' of the system were characterised by ground-truth source-to-sensor gain matrices ${\bf A}_0$, ${\bf B}_0$, ${\bf C}_0$ and  ${\bf D}_0$, calculated with the Biot-Savart equation using code from the VBMEG toolbox (vbmeg.atr.jp) \citep{takeda_meg_2019}. 
To simulate the learning phase, a 26-row driving current matrix with Gaussian-distributed values was multiplied by ${\bf C}_0$ and  ${\bf D}_0$ to obtain ${\bf x}_d (t)$ and  ${\bf r}_d (t)$, from which ${\bf C}$ and  ${\bf \Delta}$ were estimated (Methods). Similarly, a 26-row ambient source matrix was multiplied by ${\bf A}_0$ and  ${\bf B}_0$ to obtain ${\bf x}_a (t)$ and  ${\bf r}_a (t)$ from which ${\bf P}$ was estimated. Finally, ${\bf M}$ was estimated according to Eq.\ \ref{M}.

To simulate the shielding phase, a 26-row ambient source matrix was multiplied by ${\bf A}_0$ to obtain the uncorrected ambient field at the primary sensors, and by ${\bf C}_0{\bf M\Delta}{\bf B}_0$ to obtain the corrective field (${\bf A}_0$, ${\bf C}_0$ and ${\bf B}_0$ represent propagation within the ``physical'' system, the other terms are data-driven elements of the active shielding system). The corrective field was subtracted from the ambient field to get the corrected (residual) field.  The distributions of amplitude of the uncorrected and residual fields are displayed in Fig.\ \ref{simulation}B and C. Across the 274 sensors, the ambient field power is reduced by a factor of $\sim$10$^8$ (reflecting numerical limits).  

\paragraph{Additional simulations.}
\label{additional_simulations}
Using the same geometry, the ``ambient'' sources here are driven by a 28-channel recording from the reference magnetometers of a CTF system at 4800~Hz sampling rate in an ``empty room'' condition. Means were removed (they reflect the operating point of the SQUID flux lock loop), channels were normalised for equal variance, PCA was applied to render the ``sources'' decorrelated, and the PCs normalised for equal variance and used to drive ``ambient sources'' with locations as in Fig.\ \ref{simulation}A. Their amplitudes were further scaled so that the total variance of the simulated field over the primary sensor array (after application of ${\bf A}_0$) equalled that measured over the CTF reference sensor array (after mean removal) in the original recording. A subset of 25 channels were selected to drive the simulated sources (crosses in Fig.\ \ref{simulation}A), and a final ``source'' was given a constant, non-zero value to simulate the presence of a DC field.

\begin{figure}
\includegraphics[scale=.5]{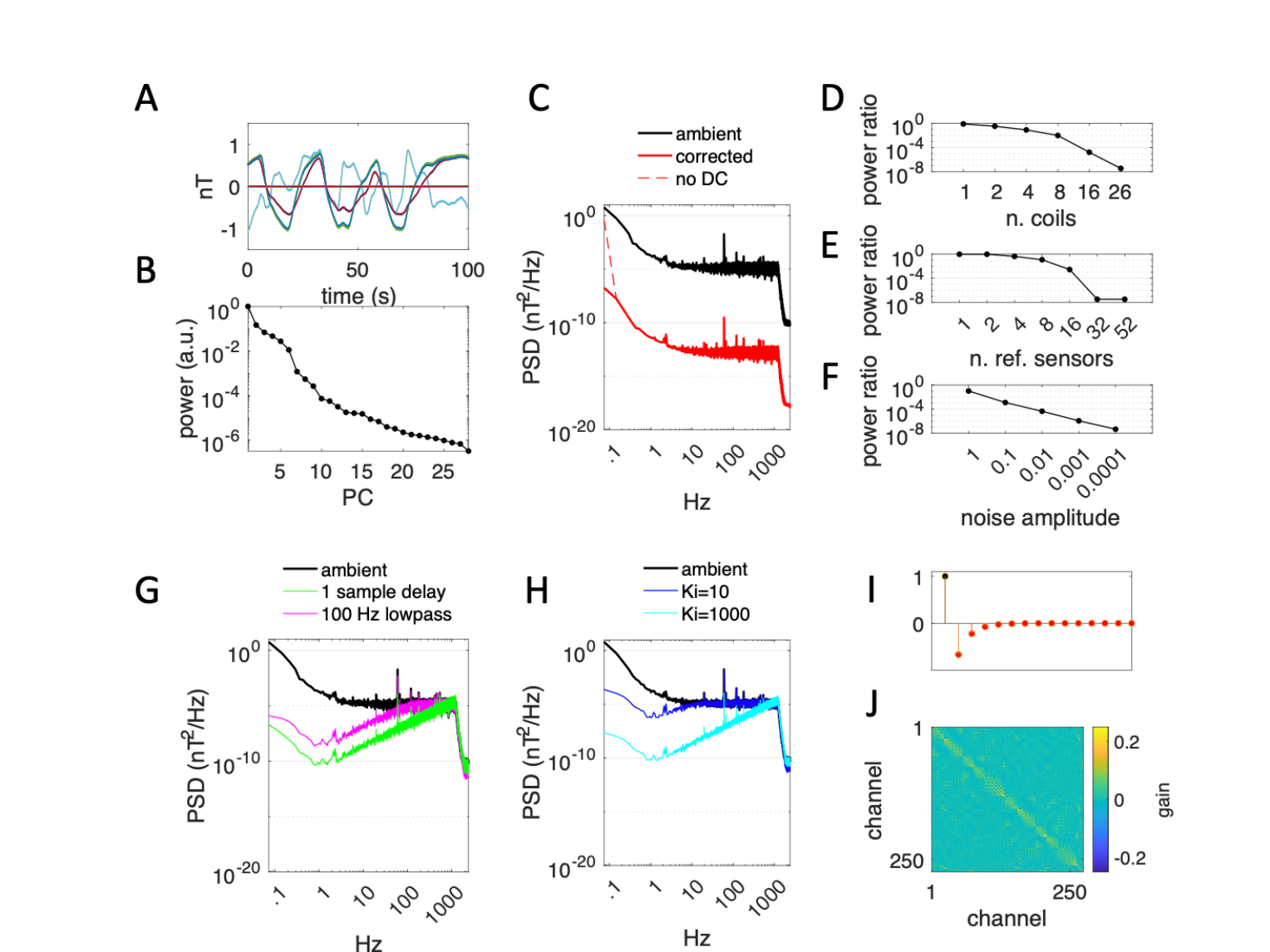}
\caption{\label{simulation2}
A: Time course of CTF reference magnetometers. 
B: PCA spectrum of normalised time courses. 
C: PSD of simulated ambient field at primary sensors (black), corrected field at primary sensors for DC-sensitive sensors (red) or DC-insensitive sensors (red dashed). 
D: Field power attenuation ratio as a function of the number of coils. 
E: Field power attenuation ratio as a function of the number of reference sensors. 
F: Field power attenuation ratio as a function of the relative amplitude of the noise. 
G: PSD of simulated ambient field at primary sensors (black), and corrected field in the hypothesis of a one-sample delay (1/4800~s) in the corrective field (green), or in the hypothesis that reference sensors have a first-order lowpass characteristic with 100~Hz cutoff (magenta). 
H: PSD of simulated ambient field (black) and PID-corrected fields for Ki=10 (blue) or Ki=1000 (cyan). 
I: Equivalent impulse response of PID controlled field with K$_i$=10 (black corresponds to direct pathway, red to control pathway). 
J: Point-spread matrix of the MIMO PID controller.
}
\end{figure}

Waveforms measured at the original CTF reference sensors are shown in Fig.\ \ref{simulation2}A, and the PCA spectrum of their normalised waveforms in panel B. The drop over PCs indicates strong correlation between sensors, but the fact that relative power does not fall below $\sim$10$^{-6}$ (10$^{-3}$ in amplitude) is consistent with at least 26 distinct ambient sources in the original environment (although one cannot exclude a contribution from sensor noise). The power spectral density (PSD) at the primary sensors is plotted in panel C (black). Elevated values near 0~Hz correspond to the slow variations visible in panel A, plausibly due to moving machinery within environment of the original recording, sharp peaks represent power line components, and the flat portion reflects the sensor noise floor (the upward trend near the antialiasing cutoff is unexplained). The leftmost value (DC) reflects the arbitrary amplitude of the ``DC source''. Spectral and correlation properties of this ``ambient field'' are more realistic than those of  the previous simulation.

The shielding solution was calculated using the first 50~s of the recorded data, and tested using the second 50~s of those data. The spectrum of the residual (panel C, red) resembles that of the ambient field (black) but with power reduced by a factor of  $\sim$10$^8$ across the entire spectrum, consistent with the previous simulation.  Sensors were assumed to have a wide bandwidth including DC, but if  the reference and/or primary sensors were unable to measure a DC field (as is the case of the SQUID sensors of the CTF system), DC components would not be attenuated (red, dashed).  Panels D, E, and F show how performance depends on the number of coils, number of reference sensors, and sensor noise variance, respectively.

In a digital implementation, the feed-forward pathway would include a delay of at least one sample, which degrades performance as illustrated in panel G (green) for a sampling rate of 4800~Hz. This effect would be alleviated running the system at a higher sampling rate. A low-pass characteristic of the reference sensors (panel G, magenta) has a similar effect, which could be alleviated by the use of  higher-bandwidth  sensors. An appealing option is to use {\em induction coils}\ \citep{brake_new_1993, tumanski_induction_2007} as reference sensors, differentiating ${\bf x}(t)$ before applying Eq.\ \ref{P} in the learning phase,  and integrating ${\bf q}(t)$ after applying Eq.\ \ref{Q} in the shielding phase. Simulation of this option yields the same result as plotted in Fig.\ \ref{simulation}C (red, dashed), with good field cancellation across the spectrum (except DC). 

The shielding system was designed with enough coils and sensors to span the ambient field (see Sects. 4.3 and 4.4 for what this entails). Performance degrades if there are fewer coils (Fig.\ \ref{simulation}D) or reference sensors Fig.\ \ref{simulation}E), or if sensor noise reduces the accuracy of the data-driven estimates of ${\bf C}$ and ${\bf P}$ (Fig.\ \ref{simulation}F). 

\paragraph{Feed-back shielding.}

For the purpose of comparison, the same configuration (Fig.\ \ref{simulation}A) was used to simulate feed-back shielding, ignoring the reference sensors and driving the coils from the primary sensors via a PID controller. The multiple interacting control pathways call for a MIMO (multiple input multiple output) PID design, which can be implemented by estimating the coil-to-primary sensor matrix ${\bf C}$ with the data-driven approach described in Methods, and using its pseudo inverse ${\bf C}^\dag$ as a decoupling matrix. This allows the control problem to be solved by $L$ independent SISO (single input single output) PID controllers.

The $K_p$ and $K_d$ parameters (proportional and differential) were set to zero (making each PID controller effectively an I controller) and the $K_i$ parameter (integral) was set to the same value for all the controllers.  The PSD of the corrected field is plotted in Fig.\ \ref{simulation}G for $K_i$=10 (blue) and $K_i$=1000 (cyan). For larger values of $K_i$ the system is unstable (as may occur also if $K_p$ and/or $K_d$ are non-zero).
Shielding is most effective in the low-frequency region.  In its basic form (no clamping), a PID controller is a linear time-invariant system that can be assimilated to a filter. The impulse response of the control pathway for $K_i$=10 is plotted in Fig.\ \ref{simulation2}I (red), together with the impulse response of the direct pathway (a Dirac pulse, black). Together, they define a first-order high-pass filter, consistent with effects shown in panel H, with a cutoff frequency that depends on $K_i$ . The slope of a first-order filter is rather shallow; a steeper slope could be achieved with a PI$^2$D controller \citep{zheng_design_2025}, and so-on.  
The PID behaves like a high-pass filter applied to the ambient field to attenuate the low-frequencies that dominate it. Small non-zero values of  $K_p$ and $K_d$ change the shape of the smoothing kernel, with little useful effect hence they are set to zero. 

As mentioned in the Introduction, a feed-back field cancelling system driven from a sensor cannot differentiate between ambient and target fields at that sensor: a target field (e.g., from the brain) triggers a corrective field that counters it. Moreover, since there are only $L<J$ control variables, there must be leakage between sensors as quantified by a point-spread matrix (${\bf C}{\bf C}^\dag$, panel J), so noise on one sensor must spread to others via the corrective field. These effects could be avoided by applying the controller to reference sensors distant from target sources, or by tuning it so that its high-pass cutoff falls below target frequencies of interest \citep{iivanainen_scalp_2019}, or by using a feed-forward design.

\section{Discussion}

Active shielding is a potentially  useful alternative or complement to classic passive shielding. Many investigators have developed methods towards this goal, and the approach described here follows in their footsteps. Simulations suggest performance competitive with other approaches, but questions remain regarding relative merits of different approaches, and  issues that might arise in their design and implementation. The next subsections point to some of these issues, and to possible paths to solve them. 

\subsection{Feed-back, feed-forward, or both?}
Feed-back shielding is necessarily imperfect, as perfect cancellation would deprive the ascending path in Fig.\ \ref{topology} (left) of input. In contrast, feed-forward shielding can be perfect in principle, but in practice, there are multiple tradeoffs.
Most proposals are of the feed-back variety \citep[e.g.][]{cohen_large-volume_1970, kelha_design_1982, brake_improvement_1991, hamalainen_m_magnetoencephalograpy_1993, vrba_magnetoencephalography_2002, boto_moving_2018, iivanainen_scalp_2019, pratt_kernel_2021, labyt_serf-opm_2022, hillebrand_non-invasive_2023, holmes_enabling_2023, jarm_active_2024, tian_active_2024,  Bezsudnova2026.08.06.743270}, which itself can be divided according to whether the control sensors are the ``primary sensors'' of a field measurement system such as MEG \citep[e.g.][]{iivanainen_scalp_2019, holmes_balanced_2019, holmes_enabling_2023}, or dedicated reference sensors within the region to be controlled \citep[e.g.][]{brake_improvement_1991, Bezsudnova2026.08.06.743270}.  Note that ``feed-back'' here refers to the ambient field control loop, as distinct from a loop internal to the sensor \citep{fourcault_helium-4_2021}.
Driving the shield from the primary sensors might seem cost-effective, but it prevents those sensors from measuring, e.g., brain fields, as they too would be cancelled, motivating the recourse to dedicated reference sensors. Alternatively, a technique such as Signal Space Separation \citep{taulu_spatiotemporal_2006} might  be used to partition between target and ambient fields, at the cost of reduced  dimensionality of the measurement data. 

Feed-back systems typically employ a small or moderate number of coils, ranging from one \citep{kelha_design_1982, brake_improvement_1991} to 
48 \citep{holmes_enabling_2023}, and a small or moderate number of reference sensors, ranging from zero (if primary sensors are used to drive the feed-back loop) \citep{iivanainen_scalp_2019} or one  \citep{kelha_design_1982, brake_improvement_1991} to 10 \citep{forte_automatic_2010}.
PID control is easy to implement for a single sensor/coil pair, and readily extended to a small number of control loops (e.g. three field axes) \citep{holmes_balanced_2019}, but a larger number of loops requires decoupling techniques \citep[e.g.][]{forte_automatic_2010} such as implemented in Sect.\ \ref{additional_simulations}.

The feed-forward approach is less common \citep{nakashima_experimental_2010, pyragius_high_2021}.
It has  potential advantages over feed-back, as mentioned in the Introduction, but to fully realise them requires overcoming spurious coil-to-reference sensor coupling (${\bf D}$), as investigated in this paper.
Recourse to a decoupling matrix is known in acoustic field control \citep{cabell_principal_1998, poshtan_multichannel_2003}, but it does not seem to have been applied to feed-forward magnetic shielding. This decoupling matrix is distinct from (but in the same simplifying spirit as)  that used to implement a MIMO PID controller.

\subsection{With, or without, a passive shield?}

Active shielding is often proposed as a complement to passive shielding \citep[e.g.][]{brake_improvement_1991, holmes_lightweight_2022}.  A passive shield can offer a very large shielding factor over most of the spectrum, apart from the low frequency region \citep{brake_improvement_1991}, but it  may introduce convolutional effects, non-linearity,  and spatial non-uniformity due to  ``mirror sources'' \citep{donoghue_planar_2015, zhao_non-uniform_2025, xu_active_2026, brake_improvement_1991, liu_enhancement_2024, zheng_design_2025} that complicate the active shield design. The designer must choose where to place reference sensors and coils relative to the MSR, e.g. all inside \citep{brake_improvement_1991}, or all outside \citep{kelha_design_1982, holmlund_novel_2000, wei_nested_2023}, or some other combination \citep{brake_new_1993, zhao_dynamic_2024}).  
The shield itself might contribute noise, either intrinsic to the shielding material \citep{liu_enhancement_2024}, or  remanent fields that require degaussing \citep{nordenstrom_feasibility_2024,schofield_novel_2024,wang_noise_2025}, or vibration-induced currents due to the earth's magnetic field \citep{volegov_noise-free_2004}. 

Doing away with the passive shield would avoid these issues and remove a major element of cost. Pleas have been made in favor of shieldless magnetocardiography (MCG)  \citep{brisinda_clinical_2023}, and multiple studies attempt to work without a passive shield \citep{aarnink_active_1995, limes_portable_2020, Bezsudnova2026.08.06.743270}.

\subsection{Active shielding, downstream processing, or both?}

Assuming linearity, subtraction of the corrective field in the field domain amounts to subtracting the corresponding signal (after analog-to-digital conversion) in the digital domain. The advantage of the former is reduced susceptibility to nonlinearities and improved dynamic range. The approaches are deeply related, and feed-forward shielding can be seen as an additional way to implement a null filter (spatial filter that cancels a specific source, in this case ambient) \citep{de_cheveigne_spatial_2026}. The two approaches might usefully be combined, for example by feeding a measure of current noise to downstream processing.



\subsection{Number, location and shape of coils}

Requirements for the coil array depend on the complexity of the ambient field over both space and time.
At time $t$, the $L$ coils must produce a corrective field at the $J$ sensors with the same distribution as the ambient field but opposite sign. This can always be achieved with $L=J$ coils if the matrix ${\bf C}$ is of full rank (i.e., coil field patterns are sufficiently diverse), but a smaller number might suffice if the ambient field pattern is simple enough.   For example $L$=3 well-designed coils might reproduce a uniform field with arbitrary orientation, $L$=8 coils a field with uniform gradients (the 9 spatial gradients reduce to 5 thanks to Maxwell's equations), and so on.  A low-order approximation may suffice if ambient sources are distant \citep{tierney_modelling_2021}, however more proximal ``mirror currents'' might complicate the field. Complex field patterns can be approximated using shimming techniques \citep[e.g.][]{turner_target_1986, kutschka_magnetic_2021, holmes_bi-planar_2018, tian_design_2025, zhao_non-uniform_2025}.

The coil array must also approximate the ambient field over all time. Applying singular value decomposition (SVD) to the ambient data matrix, ${\bf x}_a(t)={\bf U \Sigma V}$, the first $r$ columns of ${\bf U}$ span the space of spatial patterns over primary sensors, where $r$ is the rank of ${\bf x}_a(t)$.  The spatial patterns of the $L$ coils must also span that space.  To empirically check that this is the case, canonical correlation analysis (CCA) an be applied to the mixing matrix ${\bf C}$ and the first $r$ columns of ${\bf U}$. The $r$ correlation coefficients should all equal one, if not the coils must be redesigned, or the coil array  augmented with coils with additional patterns. The latter option might be easier in a data-driven context.
Thus, the number $L$ of coils should be at least $r$, rank of the ambient field. The rank $r$ might be relatively small ($<<J$) if ambient sources sources are few and/or linearly dependent; for example the steep PCA spectrum of Fig.\ \ref{simulation2}B suggests that the ambient field was well captured by a handful of PCs. Regardless of this argument, $L>r$ might  be required to capture a complex spatial pattern unless the coils are specifically designed to approximate it. 
These considerations pertain to the {\em ideal}\ number of coils, but a smaller number might yield an acceptable approximation, as illustrated in Fig.\ \ref{simulation2}D,
Note that the same constraints apply for  {\em any}\ active field control system that aims at perfect field cancellation. They are not specific to the scheme proposed here.

\subsection{Number, location and type of reference sensors}

The reference sensors must fully span the ambient field. This might not be true if (a) the reference sensors miss some dimension of that field (e.g., if $K<r$) or (b) if the reference sensors are affected by additional ``nuisance sources.'' To empirically check if reference sensors are adequate, CCA can be applied to ${\bf r}_a(t)$ and the $r$ columns of matrix ${\bf V}$ from the SVD of ${\bf x}(t)$. The $r$ correlation coefficients should be all ones, if not, the reference array must be augmented if we wish  to achieve accurate prediction of ${\bf x}(t)$ from ${\bf r}(t)$.  It must be  further augmented by $L$ reference sensors to implement the decoupling matrix, although the need for these additional sensors might be waived if coil to reference sensor coupling were minimised, for example by using coils internal to the primary sensors \citep{fourcault_helium-4_2021, tayler_miniature_2022}.

Again, these constraints pertain to the {\em ideal}\ number of reference sensors, but a smaller number might yield an acceptable approximation, as illustrated in Fig.\ \ref{simulation}E. Reference sensors should be placed so that they do not pick up target fields (e.g., from the brain) \citep{weinstock_multichannel_2000}. Sensor sensitivity, dynamic range, linearity, and noise (see below) are important factors, but reference sensors are not required to pick up tiny fields.  
Similar requirements (apart from the extra reference sensors for the decoupling matrix) apply  for feed-back shielding, to ensure that zeroing the ambient field at the reference sensors also zeros it at the primary sensors. 
 
\subsection{Dynamic range, noise, and bandwidth}

Production of a corrective field of moderate magnitude could involve the superposition of large positive and negative terms, requiring unreasonably large currents. This can be quantified by an ``inefficiency index'' defined as  the average (or maximum) over $t, j$ of $\sum_l \lvert c_{lj}q_l(t) \rvert / \lvert \sum_l c_{lj}q_l(t) \rvert$. This index  (the smaller the better) depends on both ${\bf C}$ and ${\bf q}(t)$, and thus it might be reduced by a  different design for the coil array, or by choosing $L>r$ so that there exist multiple solutions for ${\bf q}(t)$ to choose among.  Details of how this might be done are outside the scope of this paper. 

Noise can originate in reference sensors, preamplifiers, current amplifiers and coils \citep{durdaut_low-frequency_2021}. To the extent that noise is independent between coils, field noise will be multiplied by the ``inefficiency'' index just mentioned.  Current noise might reduced by reducing that index, or removed  by measuring the actual currents and feeding the difference from intended currents ${\bf q}(t)$ to downstream processing, at the cost of additional ADC channels.

The bandwidth is determined by the equivalent filter comprised of the direct pathway (e.g. black in Fig.\ \ref{simulation2}I) and corrective pathway (red).  Main factors are the  delay of one path relative to the other (which can be reduced by running at a higher sampling or buffer rate or using a programmable analog gain matrix, \citep{schlottmann_highly_2011}), and the bandwidth of the reference sensors. These considerations (current limits, noise, and to some extent bandwidth) apply to any active shielding scheme.

An option worth investigating is that of ``nested active shields'' \citep{wei_nested_2023}, a first simple shield taking care of the DC field and/or uniform field components using high-field sensors, and a second shield taking care of remaining non-uniform and/or rapidly-varying field components, using lower-field sensors. Turning them on in sequence would allow the system to be ``bootstrapped'', and the second system could suppress noise introduced by the first (its coils being seen as ``sources'' by the second system).

If  coils were placed inside an MSR and reference sensors outside, the corrective pathway  would lack the MSR-induced filtering undergone by the direct pathway. This can be addressed by augmenting the reference data with time shifts (or other convolutional basis) to allow the predictive matrix to implement a convolutional transform \citep{de_cheveigne_denoising_2007}.

\subsection{Cost}

Cost is in a tradeoff with performance, modulated by the available technology. To drive $L$ coils requires the same number of low-noise current sources and DAC channels as for a classic feed-back design with similar performance goals, but the number of reference sensors required here is greater because of the decoupling matrix.  
This extra cost is offset by ease of design of a decoupled system, and expected performance benefits.
While considerable, the cost not incommensurable with that of an MEG system: a manufacturer might find that a high-performance active shield is a cost-effective sales argument.  
Induction coils (Sect.\ \ref{additional_simulations}) might be worth considering as a cheaper alternative to magnetometers \citep{holmlund_novel_2000, pellicer-guridi_towards_2017, lin_review_2022} for reference sensors.

\subsection{A wider application space}

This paper focused on field control at the primary sensors of a measurement system such as MEG with fixed geometry.  Other applications (e.g. medical, scientific or industrial) might require control within a {\em volume}. Examples are
zero or ultra-low field (ZULF) spectroscopy \citep{blanchard_lower_2021}, quantum computing, or ultra-sensitive magnetic field metrology.  This might be achieved by surrounding the volume with primary sensors, taking into account the field complexity constraints mentioned above.  Yet other applications (e.g. ambulatory MEG) might call for a {\em variable geometry} such as a subject moving in space with a sensor array on their head  \citep[][]{rea_precision_2021, holmes_enabling_2023,boto_moving_2018}. 
One approach is to control the field within a wide enclosing volume, another is to adjust the shielding solution adaptively; both might benefit from the present ideas. 
The decoupling-matrix approach might also be applicable to other situations where transducers (e.g. electric or acoustic) impinge on sensors that control them, an example being 
the suppression of artifactual electric fields due to stimulation in brain-computer interface or medical applications.

\subsection{In sum}
This paper described an approach to  magnetic shielding that may usefully complement other approaches. Theory suggests that the residual field might be reduced to zero, but approximating that goal requires overcoming several issues, such as the choice of geometry and position of coils and sensors, current limits, and noise, some of which are shared by any active shielding scheme. Speculative ``solutions'' were considered but not simulated or implemented, the aim being to highlight potential strategies rather than claim their efficacy.

\section*{Conclusion}

Ambient magnetic field suppression is of critical importance for accurate  field measurements. This study suggests that a feed-forward approach to active shielding might lead to high quality ambient field suppression, as long as spurious coil-sensor coupling is dealt with, which can be done algorithmically at the cost of additional sensors. The method performed well in simulations, setting the stage for future implementations in hardware.

\section*{Data and code availability}
Code for figures 4, 5, 7, 8 is available on GitHub.  

\section*{Author contribution}
All aspects of this paper are under the responsibility of the author.

\section*{Funding}
This work was partially supported in part by  FrontCog grant ANR-17-EURE-0017 

\section*{Competing interests}
none

\section*{Ethics}
n/a

\section*{Acknowledgements}
 The paper benefitted from useful comments from Jacob Reichel, Tilmann Sander-Thoemmes, and Gen Uehara.

{\setstretch{1}

\end{document}